\documentclass[showpacs,aps,prd,reprint,superscriptaddress,nofootinbib,longbibliography]{revtex4-2}
\usepackage[colorlinks=true, pdfstartview=FitV,
bookmarks=true, bookmarksnumbered=true, breaklinks]{hyperref}
\usepackage[dvipdfmx]{graphicx}
\usepackage{amsmath,amssymb,bm,color,longtable,mathrsfs,slashed,comment,tikz,enumerate}

\definecolor{darkviolet}{rgb}{0.58, 0.0, 0.83}
\definecolor{electricultramarine}{rgb}{0.25, 0.0, 1.0}
\definecolor{brightpink}{rgb}{1.0, 0.0, 0.5}
\hypersetup{linkcolor=brightpink, citecolor=darkviolet, urlcolor=electricultramarine}

\definecolor{lime}{HTML}{A6CE39}
\DeclareRobustCommand{\orcidicon}{
	\hspace{-3mm}
	\begin{tikzpicture}
	\draw[lime, fill=lime] (0,0) 
	circle [radius=0.16] 
	node[white] {{\fontfamily{qag}\selectfont \tiny ID}};
	\draw[white, fill=white] (-0.0625,0.095) 
	circle [radius=0.007];
	\end{tikzpicture}
	\hspace{-3mm}
}

\foreach \x in {A, ..., Z}{\expandafter\xdef\csname orcid\x\endcsname{\noexpand\href{https://orcid.org/\csname orcidauthor\x\endcsname}
			{\noexpand\orcidicon}}
}

\begin{document}

\title{Jordan-Wigner mapping between quantum-spin and fermionic Casimir effects}

\author{Katsumasa~Nakayama\orcidB{}}
\email[]{katsumasa.nakayama@riken.jp}
\affiliation{RIKEN Center for Computational Science, Kobe, 650-0047, Japan}

\author{Kei~Suzuki\orcidC{}}
\email[]{k.suzuki.2010@th.phys.titech.ac.jp}
\affiliation{Advanced Science Research Center, Japan Atomic Energy Agency (JAEA), Tokai, 319-1195, Japan}

\begin{abstract}
The Jordan-Wigner transformation connects spin operators in one-dimensional spin systems and fermionic operators. 
In this work, we elucidate the relationship between the finite-size corrections in the spin representation and the fermionic Casimir effect in the corresponding fermion representation.
In particular, we focus on the ground-state energy of one-dimensional transverse-field Ising and XY models, and show that all finite-size corrections can be interpreted as lattice fermionic Casimir effects.
We further find several types of Casimir phenomena, such as the conventional Casimir energy from massless fields, damping behavior from massive fields, vanishing behavior from flat or nonrelativistic bands, and oscillating behavior from the finite-density effect.
Our findings establish a dictionary between finite-size corrections in spin chains and fermionic Casimir effects, and provide experimentally relevant platforms for the fermionic Casimir phenomena.
\end{abstract}

\maketitle

\section{Introduction} \label{Sec:1}

The Casimir effect has been well known theoretically~\cite{Casimir:1948dh} and experimentally~\cite{Lamoreaux:1996wh,Bressi:2002fr} as measurable forces between conducting plates (see Refs.~\cite{Plunien:1986ca,Mostepanenko:1988bs,Bordag:2001qi,Milton:2001yy,Klimchitskaya:2009cw,Woods:2015pla,Gong:2020ttb,Lu:2021jvu} for reviews).
While the conventional Casimir effect originates from quantum vacuum fluctuations of photon fields, analogous (or modified) phenomena occur also in fermion fields~\cite{Johnson:1975zp,Mamaev:1980jn}, lattice spacetime~\cite{Actor:1999nb,Pawellek:2013sda,Ishikawa:2020ezm,Ishikawa:2020icy}, spin systems~\cite{Blote:1986}, and critical systems~\cite{Fisher:1978}.\footnote{The critical Casimir effect~\cite{Fisher:1978} (see Ref.~\cite{Garcia:1999} for experiments and Ref.~\cite{Dantchev:2022hvy} for a review) is well known as a thermodynamic analog.
In this context, the Casimir effect in one-dimensional classical spin chains is well-defined~\cite{Rudnick:2010}.
Our analysis using the JW transformation provides a comprehensive understanding of classical and quantum spin systems.}

In this work, we focus on finite-size corrections to the ground-state energy in one-dimensional spin systems (i.e., spin chains) that can be mapped to fermionic systems via the Jordan-Wigner (JW) transformation~\cite{Jordan:1928wi}, and interpret them as Casimir effects.
Our goals are summarized as follows:\footnote{
While previous studies have used the Jordan-Wigner transformation
to investigate specific Casimir phenomena~\cite{Gonzalez-Cabrera:2010,Fouokeng:2019,Mula:2020udv},
our motivation is more general and aims at establishing a unified
framework for understanding Casimir effects in quantum spin chains.}
\begin{enumerate}
\item \textbf{Definition:} To define Casimir energies consistently with the Jordan-Wigner mapping between spin and fermion chains.
\item \textbf{Interpretation:} To interpret finite-size corrections in quantum spin chains as Casimir effects.
\item \textbf{Dictionary:} To establish a dictionary of different types of Casimir effects.
\item \textbf{Platform:} To propose realistic platforms for observing fermionic Casimir effects.
\end{enumerate}

For the first point, this is conceptually nontrivial because, in general, a Casimir energy admits multiple theoretical definitions, and it is not clear which one is suitable for interpreting finite-size corrections of spin systems as a Casimir effect.

For the second point, as an example, Bl\"ote, Cardy, and Nightingale~\cite{Blote:1986} pointed out that the finite-size scaling of the critical Ising chain can be interpreted as a Casimir effect. This interpretation is frequently mentioned in textbooks on conformal field theory (CFT), where the coefficients of first-order finite-size corrections are predicted by CFT.
While this is a striking connection between spin chains and the Casimir effect, it does not capture the full picture of finite-size scaling: in this work, we ask whether (i) the full-order finite-size corrections and (ii) the regimes beyond the CFT description can also be interpreted in terms of a Casimir effect.

To this end, we point out that fermionic Casimir effects realized in spin-chain systems are, precisely speaking, {\it lattice fermionic Casimir effects}.
The conventional Casimir effect originates from quantum fields in continuum spacetime, while the Casimir effect appears even from quantum fields defined {\it on the lattice}~\cite{Actor:1999nb,Pawellek:2013sda,Ishikawa:2020ezm,Ishikawa:2020icy,Nakayama:2022ild,Nakata:2022pen,Mandlecha:2022cll,Nakayama:2022fvh,Swingle:2022vie,Nakata:2023keh,Flores:2023whr,Nakayama:2023zvm,Beenakker:2024yhq,Fujii:2024fzy,Fujii:2024ixq,Fujii:2024woy}.
Among them, the Casimir effect from relativistic lattice fermion fields was first derived in Refs.~\cite{Ishikawa:2020ezm,Ishikawa:2020icy}.
We propose that all finite-size corrections of a spin chain can be mapped to a lattice fermionic Casimir effect.
On this basis, we can establish a dictionary of various types of Casimir effects.

Finally, regarding the fourth point, the fermionic Casimir effect~\cite{Johnson:1975zp,Mamaev:1980jn} (e.g., from electron and quark fields) has a long history but has not been explored in experimentally realizable systems.
If a spin chain can be mapped to fermionic systems via the JW transformation~\cite{Jordan:1928wi}, it provides an experimentally relevant platform for investigating fermionic Casimir phenomena.

This paper is organized as follows.
In Sec.~\ref{Sec:2}, we formulate the definition of Casimir energy in transverse-field Ising chains and discuss typical behavior from numerical results.
This definition is also applied to the transverse-field XY chains in Sec.~\ref{Sec:3}.
Section~\ref{Sec:4} is devoted to our conclusion.

We note that all numerical results presented in this paper are confirmed by comparisons with known analytic solutions as well as by exact diagonalization and density-matrix renormalization group (DMRG) calculations~\cite{White:1992zz,White:1993zza}.
We find that the Casimir energies calculated by two representations connected through the JW transformation (i.e., spin and fermion chains) are equivalent.
For the details of numerical methods, see Appendix~\ref{App:numerical}.

\section{Transverse-field Ising chain} \label{Sec:2}
The Hamiltonian of one-dimensional transverse-field Ising (TFI) model is~\cite{Pfeuty:1970qrn,Burkhardt:1985bqo,Cabrera:1986,Cabrera:1987}
\begin{align}
H_\mathrm{TFI} = -2J \sum_{i} S_i^x S_{i+1}^x - h \sum_{i} S_i^z, \label{eq:TFIM}
\end{align}
where the spin operators are represented as $S_i^x = \frac{\sigma^x}{2}$ and $S_i^z = \frac{\sigma^z}{2}$ with the Pauli matrices on the $i$ th site.
$J$ and $h$ are the coupling constant and the transverse field applied along the $z$ direction, respectively.
In this paper, we fix a ferromagnetic coupling $J>0$ (the calculation with $J<0$ is also straightforward).

\subsection{Thermodynamic limit}
\begin{figure}[tb!]
    \centering
    \begin{minipage}[t]{1.0\columnwidth}
    \includegraphics[clip,width=1.0\columnwidth]{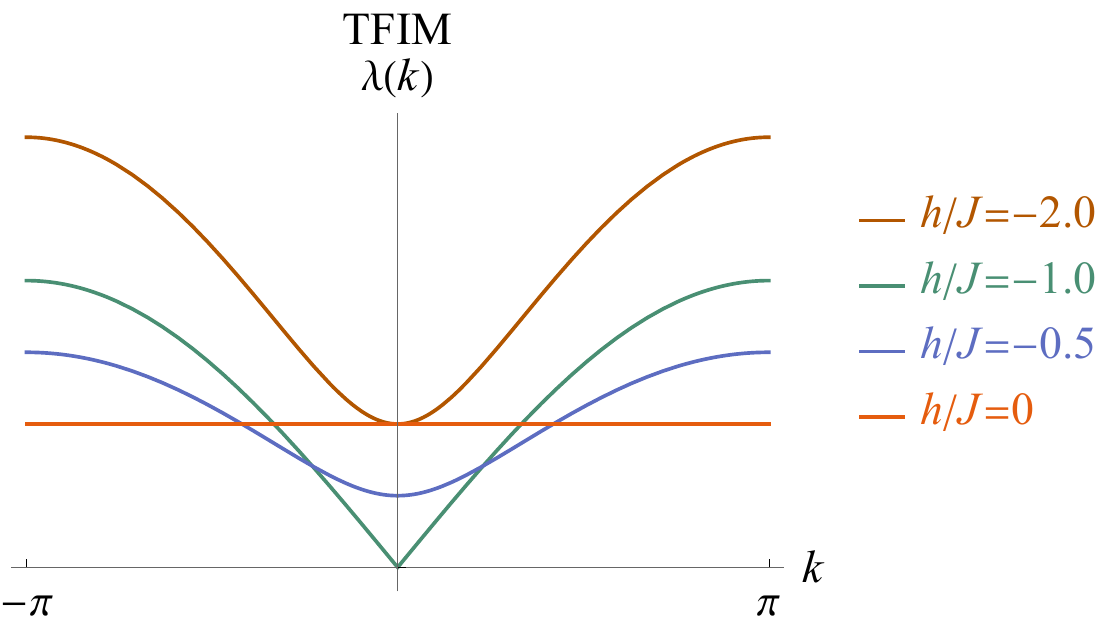}
    \end{minipage}
  \caption{Dispersion relations of fermions in the TFIM under some transverse fields $h$, based on Eq.~(\ref{eq:TFIM_lambda_bulk}).
  The dispersion relations at $h>0$ are shifted by $\pi$, compared with those for $h<0$.}
    \label{fig:TFIM_disp}
\end{figure}

The free energy at nonzero temperature $k_BT=1/\beta$ is~\cite{Pfeuty:1970qrn}
\begin{align}
F&= -\frac{N}{\beta} \left[ \ln 2 + \frac{1}{\pi} \int_0^\pi dk \ln \cosh\left( \frac{\beta}{2} \lambda_k \right) \right] \\
&= -\frac{N}{\beta \pi} \int_0^\pi dk \ln \left( e^{\frac{\beta}{2} |\lambda_k|} +  e^{-\frac{\beta}{2} |\lambda_k|}\right),  \label{eq:TFIM_F}
\end{align}
where the fermion eigenvalue is~\cite{Pfeuty:1970qrn}
\begin{align}
\lambda_k =\sqrt{J^2 +h^2 + 2Jh\cos k}. \label{eq:TFIM_lambda_bulk}
\end{align}
It is instructive to check the $k$ dependence of the eigenvalue.
In Fig.~\ref{fig:TFIM_disp}, we show typical forms of dispersion relations at $h/J=0,-0.5,-1.0,-2.0$.\footnote{Also in the case of $h/J>0$, the dispersion relations are similar but shifted by $\pi$.}
We can find characteristic behaviors such as the flat band at $h/J=0$, the gapped dispersion at $h/J=-0.5$, and the linear dispersion at $h/J=-1.0$.

From the zero-temperature limit of Eq.~(\ref{eq:TFIM_F}), the zero-point or ground-state energy of the infinite chain (i.e., ``bulk"), $E_0^\mathrm{bulk}  \equiv \lim_{T\to 0} F$, is~\cite{Pfeuty:1970qrn} 
\begin{align}
E_0^\mathrm{bulk} =   -\frac{N}{\beta \pi} \int_0^\pi dk \ln e^{\frac{\beta}{2} |\lambda_k|} =  -\frac{N}{\pi} \int_0^\pi dk \frac{1}{2} |\lambda_k|.
\label{eq:TFIM_E0bulk}
\end{align}

\subsection{Open boundary}
For the open (or free) boundary condition (OBC), the Hamiltonian (\ref{eq:TFIM}) is modified as~\cite{Pfeuty:1970qrn}
\begin{align}
H_\mathrm{TFI}^\mathrm{OBC} = -2J \sum_{i=1}^{N-1} S_i^x S_{i+1}^x - h \sum_{i=1}^{N} S_i^z. \label{eq:TFIM-OBC}
\end{align}
Thus, the OBC is imposed as a truncation of spins in the first term, using $\sum_{i=1}^{N-1}$.
The JW transformation~\cite{Jordan:1928wi} maps a Hamiltonian expressed in spin operators to the form written in the fermionic annihilation and creation operators, $c$ and $c^\dagger$, respectively:
\begin{align}
\sigma_i^x &= (c_i^\dagger + c_i) \exp \left( i\pi \sum_{j=1}^{i-1} c_j^\dagger c_j \right), \\
\sigma_i^y &= -i(c_i^\dagger - c_i) \exp \left( i\pi \sum_{j=1}^{i-1} c_j^\dagger c_j \right), \\
\sigma_i^z &= 2 c_i^\dagger c_i - 1.
\end{align}
Using this transformation, the Hamiltonian in terms of pseudo-fermions in position space is written as~\cite{Pfeuty:1970qrn}\footnote{Note that, in Ref.~\cite{Burkhardt:1985bqo}, the the sign in the third equation of JW transformation is opposite: $\sigma_i^z = 1- 2 c_i^\dagger c_i$.
As a result, the role of $h$ is effectively replaced by $-h$ in the fermion representation.}
\begin{align}
H =& -\frac{J}{2} \sum_{i=1}^{N-1} \left( c_i^\dagger c_{i+1} - c_i c_{i+1}^\dagger + c_i^\dagger c_{i+1}^\dagger - c_i c_{i+1} \right) \notag\\
&-h\sum_{i=1}^{N} (c_i^\dagger c_i -\frac{1}{2}).
\end{align}
The fermion eigenvalues are derived from the diagonalized Hamiltonian after the Fourier and Bogoliubov transformations~\cite{Pfeuty:1970qrn}:
\begin{align}
\lambda_\theta =\sqrt{J^2 +h^2 + 2Jh\cos \theta}, \label{eq:TFIM_lambda}
\end{align}
This form is the same as Eq.~(\ref{eq:TFIM_lambda_bulk}), but $\theta$ is a discrete parameter satisfying the following equation (i.e., discrete-mode condition)~\cite{Pfeuty:1970qrn} 
\begin{align}
\frac{\sin(N+1) \theta}{\sin N\theta} = -\frac{J}{h}. \label{eq:TFIM_sin/sin}
\end{align}
This equation can be numerically solved, whereas for a specific case, we have an analytic solution.
Note that the number of mathematical solutions is $2N$, while that of independent degrees of freedom is $N$.
The solutions appear as pairs of positive and negative eigenvalues, where only the positive-energy modes correspond to independent physical quasiparticle excitations.
For this reason, to calculate the zero-point energy, we pick up only the positive eigenvalues.

The zero-point energy is calculated by summing over the $N$ eigenvalues.
\begin{align}
E_0 = - \frac{1}{2} \sum_{m=0}^{N-1} \lambda_\theta, \label{eq:TFIM_E0open}
\end{align}
where $m$ is the label of discrete parameter.
This form can be regarded as the zero-point energy of $N$ Majorana fermions, and at the same time, it is equal to one-half of the zero-point energy of the corresponding Dirac fermions.

\subsubsection{Zero field: $h/J=0$}
When the magnetic field is absent (i.e., the classical Ising chain at $h/J=0$), both the eigenvalue $\lambda_\theta$ in the finite chain and eigenvalues $\lambda_k$ in the infinite chain are constant,
\begin{align}
\lambda_\theta= \lambda_k  = J.
\end{align}
For the infinite chain, by substituting $\lambda_k  = J$ into Eq.~(\ref{eq:TFIM_E0bulk}), the zero-point energy is
\begin{align}
\frac{E_0^\mathrm{bulk}}{J} &= -\frac{N}{2}, \label{eq:E0bulk_claIsing}
\end{align}
where $N$ is interpreted as the trivial factor proportional to its size: the energy density per size, $E_0^\mathrm{bulk}/N$, is a constant independent of $N$.
For the finite chain, using the classical Ising chain Hamiltonian $-\frac{J}{2} \sum_{i=1}^{N-1} \sigma_i^x \sigma_{i+1}^x$, the zero-point energy is
\begin{align}
\frac{E_0}{J} &= -\frac{N-1}{2}. \label{eq:E0_claIsing}
\end{align}

The Casimir energy is defined as the difference between the two zero-point energies:
\begin{align}
E_\mathrm{Cas} \equiv E_0 - E_0^\mathrm{bulk}. \label{eq:ECas_def}
\end{align}
Note that, in the OBC, $E_0$ includes the surface term, so that the Casimir energy is defined as the form including the surface energy.
Thus, this definition includes constant terms as well as the finite-$N$ contribution.
Using Eqs.~(\ref{eq:E0bulk_claIsing}) and (\ref{eq:E0_claIsing}),
\begin{align}
\frac{E_\mathrm{Cas}}{J} = \frac{1}{2}.
\end{align}
Thus, we can conclude that the Casimir energy is a constant and independent of $N$ (although $E_0$ and $E_0^\mathrm{bulk}$ are $N$-dependent).
The constant $\frac{1}{2}$ is equivalent to the surface energy.
Since a constant energy offset has no physical consequences, such a Casimir energy is not physically meaningful.\footnote{The disappearance of Casimir effect was also recognized in the context of the critical Casimir effect~\cite{Rudnick:2010}, where the authors confirmed the zero-temperature limit of the classical Ising chain at finite temperature.}

This feature is often known also in the context of Casimir effect in quantum field theory. 
As shown in Fig.~\ref{fig:TFIM_disp}, the fermion eigenvalue is interpreted as a ``flat band."
In general, the Casimir energy in flat-band systems is exactly zero because the eigenenergy is independent of the particle momentum, and the momentum discretization by a change of the system size does not shift the zero-point energy.

\subsubsection{Weak-field region: $0<|h|/J<1$}

As shown in the previous section, in the limit of $h \to 0$, the Casimir energy is equivalent to the surface energy, $E_\mathrm{Cas} \to \frac{J}{2}$.
With increasing $|h|$, the Casimir energy is composed of not only the positive and $N$-independent surface energy but also a negative $N$ dependence.

\begin{figure}[tb!]
    \centering
    \begin{minipage}[t]{1.0\columnwidth}
    \includegraphics[clip,width=1.0\columnwidth]{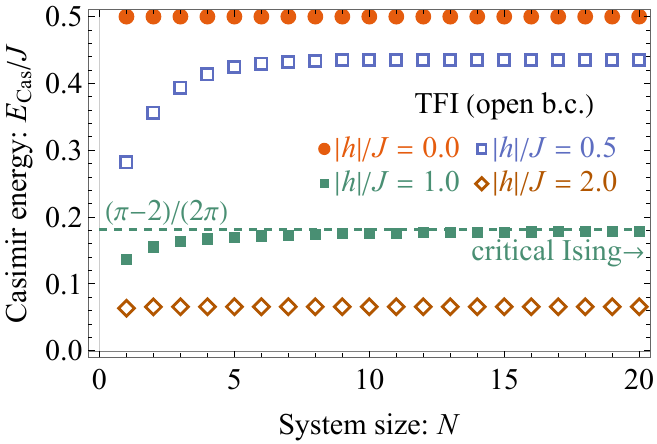}
    \end{minipage}
\caption{Casimir energy in the transverse-field Ising (TFI) chain under the open boundary condition.}
\label{fig:TFI_open}
\end{figure}

The result at $h=0.5$ is shown in Fig.~\ref{fig:TFI_open}.
We can find that the Casimir energy is suppressed in the small-size region, which corresponds to the generation of a negative Casimir energy or an attractive force.
In the large-size limit, the Casimir energy approaches a constant.
This behavior can be interpreted in terms of the dispersion relations of pseudo-fermions.
As shown in Fig.~\ref{fig:TFIM_disp}, pseudo-fermions are regarded as massive degrees of freedom with a nonzero mass gap.
In general, the Casimir energy of massive degrees of freedom is suppressed in the long range~\cite{Bender:1976wb,Hays:1979bc,Ambjorn:1981xw}.

Here, we comment on the treatment of complex eigenvalues in the pseudo-fermion representation.
In the region of $0<|h|/J<1$, it is known that some of the eigenvalues can be complex~\cite{Pfeuty:1970qrn}.\footnote{
The appearance of complex eigenvalues can be understood by the discrete-mode condition~(\ref{eq:TFIM_sin/sin}), where the left-hand side represents the ratio of wave-function amplitudes at two adjacent sites.
For certain solutions in the region of $0<|h|/J<1$, this ratio cannot be realized only with a real wave number.
Consequently, these solutions necessarily correspond to complex wave numbers, producing exponentially varying amplitudes, while other solutions remain real.}
However, when we calculate the zero-point energy by summing over all eigenvalues, the imaginary parts cancel out, and we finally obtain a real value.
Thus, the Casimir energy includes contributions from all modes, including those with complex eigenvalues, but the final value is real.
Note that this discussion is based on the eigenvalues expressed in terms of Bogoliubov-transformed pseudo-fermions.
When we perform exact diagonalization of the original spin Hamiltonian, the Hamiltonian is Hermitian, so that all the eigenvalues are real.

\subsubsection{Critical point: $|h|/J=1$}
This model at the critical point $|h|/J=1$ is known to be described as the conformal field theory (CFT) with the central charge $c=1/2$.
In this case, we can analytically understand the eigenvalue structure and the zero-point energy.

At the critical points $h=\pm J$, the discrete-mode condition~(\ref{eq:TFIM_sin/sin}) is reduced to
\begin{align}
\frac{\sin(N+1) \theta}{\sin N\theta} = \mp 1.
\end{align}
The root of this equation is known as~\cite{Burkhardt:1985bqo}
\begin{align}
|\theta| &=
\begin{cases}
\displaystyle \frac{2m+2}{2N+1}\,\pi \ \ (m=0,1,\dots,N-1) & \text{for } h=+J \\
\displaystyle \frac{2m+1}{2N+1}\,\pi \ \ (m=0,1,\dots,N-1) & \text{for } h=-J
\end{cases},
\end{align}

By substituting $h=\pm J$ into Eq.~(\ref{eq:TFIM_lambda}), the eigenvalues are
\begin{align}
\lambda_\theta &=
\begin{cases}
 J\sqrt{2+ 2 \cos \theta} = 2J \left| \cos \frac{\theta}{2} \right| & \text{for } h=+J \\
J\sqrt{2- 2 \cos \theta} = 2J \left| \sin \frac{\theta}{2} \right| & \text{for } h=-J
\end{cases}.
\end{align}

Using this form, the zero-point energy (independent of the sign of $h$) is obtained as~\cite{Burkhardt:1985bqo}
\begin{align}
\frac{E_0}{J} &= - \frac{1}{2J} \sum_{m=0}^{N-1} \lambda_\theta = \frac{1}{2}\left( 1-\csc \frac{\pi}{2(2N+1)} \right) \\
&= -\frac{2}{\pi} N +\frac{\pi-2}{2\pi} -\frac{\pi}{48N}  +\frac{\pi}{96N^2} + \mathcal{O} (1/N^3).
\end{align}
The first and second terms in the last form are the bulk contribution and the surface energy, respectively.
The third term proportional to $1/N$ is well known in the context of the CFT, which includes the universal factor $-1/24$ and the central charge $c=1/2$~\cite{Blote:1986}.

By subtracting the bulk contribution from this zero-point energy the Casimir energy is 
\begin{align}
\frac{E_\mathrm{Cas}}{J} &= \frac{1}{2}\left( 1-\csc \frac{\pi}{2(2N+1)} \right) + \frac{2}{\pi} N \\
&= \frac{\pi-2}{2\pi} -\frac{\pi}{48N}  +\frac{\pi}{96N^2} + \mathcal{O} (1/N^3).
\end{align}
Thus, the dominant contribution in the short range is the term proportional to $1/N$, as well as a constant shift $(\pi-2)/2\pi$ due to the surface energy.
The factor $- \pi/48$ is known to appear in the Casimir energy from the {\it continuous} spinless-massless Majorana fermion under the MIT-bag boundary condition.
Its minus sign is due to the fermion statistics (i.e., the minus sign of fermion zero-point energy) and physically corresponds to an attractive force.\footnote{It is instructive to consider the other well-known Casimir effect.
As a reference, for the $1+1$-dimensional real-scalar field under the Dirichlet boundary condition, we know the factor $-\pi/24$.
A factor of $-1/2$ appears due to the difference between the eigenvalue structures of Dirichlet and MIT-bag boundary conditions, and additionally a minus sign appears due to the difference between the boson and fermion statistics.
Therefore, $- \pi/48$ is also interpreted as $-\pi/24 \times (-1/2) \times (-1)$.}

It was identified by Bl\"ote-Cardy-Nightingale~\cite{Blote:1986} that the $1/N$ term is a kind of Casimir energy.
In the current work, our main finding is that the form including the higher-order terms of $\mathcal{O} (1/N^2)$ is interpreted as the Casimir energy for the {\it lattice} spinless-massless Majorana fermion under the MIT-bag boundary conditions.

\subsubsection{Strong-field region: $|h|/J>1$}

Similar to the weak-field region ($0<|h|/J<1$), the Casimir energy in the strong-field region can be interpreted as the Casimir effect of massive fermions.

The result at $h=2.0$ is shown in Fig.~\ref{fig:TFI_open}.
We can find the negative Casimir energy is mostly suppressed by the fermion mass, and the total Casimir energy is dominated by the nonzero surface energy.
As a result, the $N$ dependence is negligibly small.

We comment on complex eigenvalues of fermions.
Contrary to the weak-field region, the eigenvalues in the strong-field region does not include a complex solution.
Therefore, the Casimir energy is composed of only real fermion eigenvalues.

Finally, in the strong-field limit, $|h|\to \infty$, we can find that the Casimir energy goes to zero.
This is understood as follows: the discrete-mode condition~(\ref{eq:TFIM_sin/sin}) is
\begin{align}
\frac{\sin(N+1) \theta}{\sin N\theta} \to 0,
\end{align}
and then
\begin{align}
|\theta| \to \frac{m}{N+1} \pi \ \ \ (m=0,\cdots, N-1).
\end{align}
The eigenvalues are
\begin{align}
\lambda_\theta = h +J\cos \theta +\mathcal{O}(J^2/h) .
\end{align}
This is almost the cosine band shifted by a large $|h|$.
The Casimir energy from such a shifted cosine band under the open boundary conditions is known to be zero~\cite{Nakayama:2022ild}, which is also related to the vanishing Casimir effect from the quadratic dispersion relations in the continuum theory.

\subsection{Periodic boundary}
Next, we move to the case of the periodic boundary condition (PBC).
The Hamiltonian~(\ref{eq:TFIM}) is modified as~\cite{Pfeuty:1970qrn}
\begin{align}
\begin{aligned}
H_\mathrm{TFI}^\mathrm{PBC} &= -2J \sum_{i=1}^{N} S_i^x S_{i+1}^x - h \sum_{i=1}^{N} S_i^z, \label{eq:TFIM-PBC} \\
S_{N+1}^{x,z} &= S_1^{x,z}. 
\end{aligned}
\end{align}
The PBC for spin chains is defined as the second equation, and the case with the antiperiodic boundary condition (APBC) is also straightforward.

After the JW transformation, the Hamiltonian is rewritten as~\cite{Pfeuty:1970qrn}
\begin{align}
H =& -\frac{J}{2} \sum_{i=1}^{N-1} \left( c_i^\dagger c_{i+1} - c_i c_{i+1}^\dagger + c_i^\dagger c_{i+1}^\dagger - c_i c_{i+1} \right) \notag\\
&+ (-1)^{N_c} \frac{J}{2} \left( c_N^\dagger c_{1} - c_N c_{1}^\dagger + c_N^\dagger c_{1}^\dagger - c_N c_{1} \right) \notag\\
&-h\sum_{i=1}^{N} (c_i^\dagger c_i -\frac{1}{2}),
\end{align}
where $N_c= \sum_{i=1}^N c_i^\dagger c_i $ is the total fermion-number operator.
The parity of this system is defined as
\begin{align}
 \exp (i\pi N_c) =  \exp \left(i\pi \sum_{i=1}^N c_i^\dagger c_i \right) = (-1)^{N_c}.
\end{align}
Because this parity operator commutes with the Hamiltonian, the Hilbert space splits into even- and odd-parity sectors, where $(-1)^{N_c}$ is a fixed number, $+1$ or $-1$.

There are two types of fermion vacua characterized by the two types of momentum sets: one is the antiperiodic-type form,
\begin{align}
\theta =\frac{2m+1}{N} \pi \ \ \ (m=0,1,\dots,N-1), \label{eq:fAPBev}
\end{align}
and the other is the periodic-type form,
\begin{align}
\theta =\frac{2m}{N} \pi \ \ \ (m=0,1,\dots,N-1). \label{eq:fPBev}
\end{align}
The ground state depends on the required conditions, such as the boundary conditions of Ising chain and the evenness or oddness of $N$ and $N_c$.

When $N$ is even, for the PBC with even $N_c$, Eq.~(\ref{eq:fAPBev}) is reduced to
\begin{align}
\theta_\mathrm{no} = \pm\frac{1}{N}\pi, \pm\frac{3}{N}\pi, \dots, \pm\frac{N-1}{N}\pi. \label{eq:theta_no}
\end{align}
We call this set the {\it no-edge solution} labeled by ``no" because there is no $0$ or $\pi$ mode.
For the PBC with odd $N_c$, Eq.~(\ref{eq:fPBev}) is reduced to
\begin{align}
\theta_{0,\pi} = 0, \pm\frac{2}{N}\pi, \pm\frac{4}{N}\pi, \dots, \pm\frac{N-2}{N}\pi, \pi.
\end{align}
We labeled this set by ``$0,\pi$" because it includes both the $0$ and $\pi$ modes.

When $N$ is odd, Eq.~(\ref{eq:fPBev}) and Eq.~(\ref{eq:fAPBev}) are reduced to 
\begin{align}
\theta_{0} &= 0, \pm\frac{2}{N}\pi, \pm\frac{4}{N}\pi, \dots, \pm\frac{N-1}{N}\pi, \\
\theta_{\pi} &= \pm \frac{1}{N}\pi, \pm\frac{3}{N}\pi, \dots, \pm\frac{N-2}{N}\pi, \pi,
\end{align}
respectively.
We labeled these sets by ``$0$" or ``$\pi$" because each set includes either the $0$ or $\pi$ mode.
In Table~\ref{tab:groundstate_theta}, we summarize the momentum sets corresponding to the ground states of the TFIM.

\begin{table}[t]
\centering
\caption{
Momentum sets corresponding to ground states in the transverse-field Ising chains with PBC or APBC, where $h/J=-1$ is assumed.
If $h/J=1$, the roles of $\theta_0$ and $\theta_\pi$ is exchaged.
Also see Ref.~\cite{Burkhardt:1985bqo,Cabrera:1987}.
}

\begin{tabular}{c c c}
\hline
Spin chain & Size $N$ & Ground state \\
\hline
PBC & even & $\theta_\mathrm{no}$ \\
 & odd & $\theta_{\pi}$ \\
\hline
APBC & even & $\theta_{0,\pi}$ \\
 & odd & $\theta_{0}$ \\
\hline
\end{tabular}
\label{tab:groundstate_theta}
\end{table}

\subsubsection{Zero field: $h/J=0$}

For the infinite chain, the zero-point energy is the same as the case with the OBCs:
\begin{align}
\frac{E_0^\mathrm{bulk}}{J} &= -\frac{N}{2},
\end{align}
For the finite chain, using the classical Ising chain Hamiltonian $-\frac{J}{2} \sum_{i=1}^{N} \sigma_i^x \sigma_{i+1}^x$, the zero-point energy is
\begin{align}
\frac{E_0}{J} &= -\frac{N}{2}.
\end{align}
As a result, the Casimir energy is
\begin{align}
E_\mathrm{Cas} \equiv E_0 - E_0^\mathrm{bulk} = 0.
\end{align}
Thus, in the case of the PBC, the Casimir energy is exactly zero.\footnote{This was also recognized in the context of the critical Casimir effect~\cite{Rudnick:2010}.}
This mechanism is similar to the case of the OBC.
We note that, within our definition, whereas the Casimir energy with the OBC is equivalent to the nonzero surface energy, that with the PBC is zero due to the vanishing surface energy.

\begin{figure}[b!]
    \centering
    \begin{minipage}[t]{1.0\columnwidth}
    \includegraphics[clip,width=1.0\columnwidth]{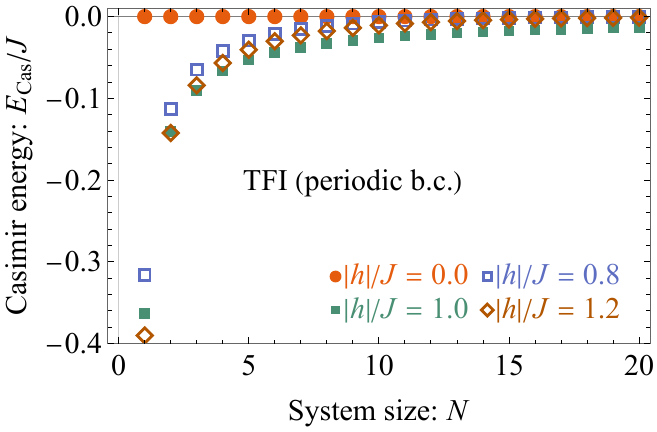}
    \end{minipage}
\caption{Casimir energy in the transverse-field Ising chain under the periodic boundary condition.}
\label{fig:TFI_pbc}
\end{figure}

\subsubsection{Weak-field region: $0<|h|/J<1$}

When a nonzero transverse field is switched on, the nonzero Casimir energy is generated.
The result at $h=0.8$ is shown in Fig.~\ref{fig:TFI_pbc}.
We find that, in the weak-field region, $0<|h|/J<1$, the Casimir energy shows a damping behavior as a function of $N$ and approaches zero in the large size.
This behavior is interpreted as a manifestation of the Casimir effect for massive degrees of freedom.
Thus, the behavior of the Casimir energy is similar to that in the case of OBC, but it differs in that no surface energy appears.

\subsubsection{Critical point: $|h|/J=1$}
In the critical points, $h/J= \pm1 $, we find that the Casimir energy scales as $1/N$ in the larger size.
This is a typical behavior of the Casimir effect for one-dimensional massless degrees of freedom.

An analytic derivation is as follows.
At the critical point in the case of PBC, the momentum sets constructing the ground state are $\theta_\mathrm{no}$ for even $N$ and $\theta_\pi$ for odd $N$.\footnote{An intuitive picture is as follows.
For even $N$, we compare $\theta_\mathrm{no}$ and $\theta_{0,\pi}$.
Using $\lambda_\theta = 2 | \sin \frac{\theta}{2}|$, the edge mode of $\theta=0$ is the lowest-energy mode among all discrete modes.
The edge mode of $\theta=\pi$ is the highest-energy mode, and the level spacing in the high-energy region is denser than that in the low-energy spectrum. 
Then, the edge mode of $\theta=0$ leads to the smallest negative energy due to the minus sign of the zero-point energy.
As a result, the ground state is $\theta_\mathrm{no}$.
Similarly, for odd $N$, the zero-point energy for $\theta_{\pi}$ is favored, compared to that for $\theta_{0}$.
}
Note that, in the case of APBC instead of PBC, the ground state is constructed from 
$\theta_{0,\pi}$ for even $N$ and $\theta_0$ for odd $N$.

As a result, the zero-point energy is (for $h/J=-1$)~\cite{Burkhardt:1985bqo}
\begin{align}
\frac{E_0^\mathrm{PB}}{J}  &= - \frac{1}{2J} \sum_{m=0}^{N-1} \lambda \left( \frac{2m+1}{N} \pi \right) =  - \csc \frac{\pi}{2N} \\
&=-\frac{2}{\pi} N -\frac{\pi}{12N}  -\frac{7\pi^3}{2880N^3}  + \mathcal{O} (1/N^5), \\
\frac{E_0^\mathrm{APB}}{J} &= - \frac{1}{2J} \sum_{m=0}^{N-1} \lambda \left( \frac{2m}{N} \pi \right)  =  - \cot \frac{\pi}{2N} \\
&=-\frac{2}{\pi} N +\frac{\pi}{6N}  +\frac{\pi^3}{360N^3}  + \mathcal{O} (1/N^5).
\end{align}
The coefficient $-\pi/12=-c\pi/6$ for the PBC is known in the context of CFT~\cite{Affleck:1986bv}.

From the definition, the Casimir energy is
\begin{align}
\frac{E_\mathrm{Cas}^\mathrm{PB}}{J} & =  \frac{2}{\pi} N - \csc \frac{\pi}{2N} \notag \\
&= -\frac{\pi}{12N}  -\frac{7\pi^3}{2880N^3}  + \mathcal{O} (1/N^5), \label{eq:EcasPB} \\
\frac{E_\mathrm{Cas}^\mathrm{APB}}{J}  &=\frac{2}{\pi} N  - \cot \frac{\pi}{2N} \notag \\
&= \frac{\pi}{6N}  +\frac{\pi^3}{360N^3}  + \mathcal{O} (1/N^5).  \label{eq:EcasAPB} 
\end{align}
The first terms, $-\pi/12N$ and $\pi/6N$, are interpreted as the Casimir energies from {\it continuous} spinless-massless Majorana fermion under the APBC and the PBC, respectively.
Their minus and positive signs are due to the fermion statistics (i.e., the minus sign of fermion zero-point energy) and physically correspond to the attractive and repulsive forces, respectively. 
Note that since a spin chain under the PBC is mapped to a fermion chain under the APBC, the above $E_\mathrm{Cas}^\mathrm{PB}$ corresponds to the fermionic Casimir effect under the APBC.

In addition, Eqs.~(\ref{eq:EcasPB}) and (\ref{eq:EcasAPB}) mean the Casimir energies from the {\it lattice} spinless-massless Majorana fermion under the APBC and the PBC, respectively.
In other words, they contain all the lattice corrections (i.e., the deviation from the continuous fermion).
These formulas (\ref{eq:EcasPB}) and (\ref{eq:EcasAPB}) coincide with those derived in Refs.~\cite{Ishikawa:2020ezm,Ishikawa:2020icy}, except the factor $1/2$ due to the difference between the Majorana and Dirac fermions.\footnote{
Note that Dirac-like fermions studied in Refs.~\cite{Ishikawa:2020ezm,Ishikawa:2020icy} are known to be the so-called Wilson fermions.
The eigenvalue in the one-dimensional massless case is proportional to $\sqrt{1-\cos{k}}$, and the low-energy dispersion relation is linear.}

\subsubsection{Strong-field region: $|h|/J>1$}
The behavior of Casimir energy in the strong-field region, $|h|/J>1$, is similar to that in the weak-field region: the Casimir energy is a monotonically decreasing function of $N$ and its damping is faster than $1/N$, as shown in Fig.~\ref{fig:TFI_pbc}.
This is interpreted as the Casimir effect for massive degrees of freedom.

In the strong field limit $|h|\to \infty$, the fermion dispersion relation can be approximated as a shifted cosine band.
Then, we find that the Casimir energy approach zero for $N>1$, which is the same result as the case of OBC.
However, only at $N=1$ (i.e., the one-site system under the PBCs), we find $E_\mathrm{Cas}/J \to 1/2$, which is distinct from the case of the OBC, where $E_\mathrm{Cas}/J \to 0$ at $N=1$.
This special behavior is known as the {\it remnant Casimir effect}~\cite{Nakayama:2022ild}, which is a feature of the Casimir effect for a shifted cosine band under the PBCs.

\subsection{Short summary}

\begin{figure}[b!]
    \centering
    \begin{minipage}[t]{1.0\columnwidth}
    \includegraphics[clip,width=1.0\columnwidth]{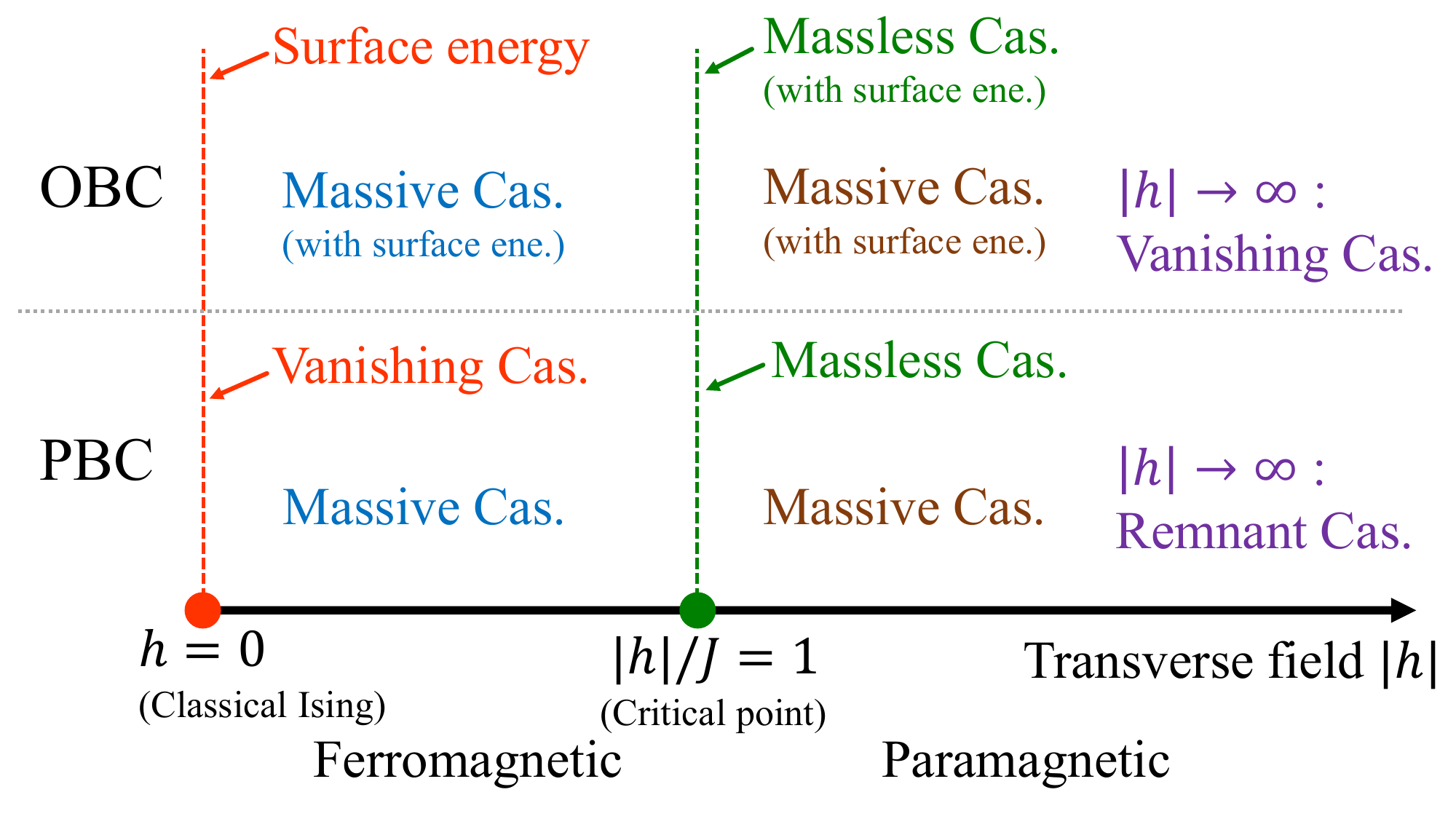}
    \end{minipage}
\caption{Phase diagram of Casimir effect in the one-dimensional transverse-field Ising model under the open or periodic boundary condition (OBC or PBC).}
\label{fig:TFI_diagram}
\end{figure}

Finally, in Fig.~\ref{fig:TFI_diagram}, we summarize a phase diagram of the Casimir effect in the one-dimensional TFIM under the OBC or PBC.
We can find various types of Casimir effect depending on transverse fields, such as the massive/massless Casimir effect and the vanishing/remnant Casimir effect.

\section{Transverse-field XY chain} \label{Sec:3}
In this section, we focus on the one-dimensional quantum XY model~\cite{Lieb:1961fr,Katsura:1962jxm,Barouch:1970ryz,Barouch:1971ywx,PhysRevA.3.2137,McCoy:1971zz}.
Although the definition of Casimir energy is the same as the case (\ref{eq:ECas_def}) of Ising chains, we can see rich and new types of Casimir effects arising from the complex phase diagram of XY model.

The Hamiltonian of the one-dimensional transverse-field XY (TFXY) model is~\cite{Katsura:1962jxm}
\begin{align}
H_\mathrm{TFXY} =& -J \sum_{i} \left[ (1+\gamma)S_i^x S_{i+1}^x +  (1-\gamma) S_i^y S_{i+1}^y \right] \notag\\
&- h \sum_{i} S_i^z,
\end{align}
where $\gamma$ is the anisotropic parameter. 
At $\gamma=1$, this Hamiltonian is reduced to the TFIM~(\ref{eq:TFIM}).
At $\gamma=0$, it is the so-called the transverse-field XX (TFXX) model:
\begin{align}
H_\mathrm{TFXX} = -J \sum_{i} \left[S_i^x S_{i+1}^x + S_i^y S_{i+1}^y \right] - h \sum_{i} S_i^z.
\end{align}

\subsection{Thermodynamic limit}

After the JW transformation, the eigenvalue is
\begin{align}
\lambda_k =\sqrt{ (J\cos{k} +h)^2 + (\gamma J \sin{k})^2}. \label{eq:TFXY_lambda_bulk}
\end{align}
At $\gamma=1$, this form is reduced to the eigenvalue (\ref{eq:TFIM_lambda_bulk}) of the TFI chain.
In the TFXX chain at $\gamma=0$,
\begin{align}
\lambda_k =|J\cos{k} +h|.
\end{align}

\begin{figure}[tb!]
    \centering
    \begin{minipage}[t]{1.0\columnwidth} \includegraphics[clip,width=1.0\columnwidth]{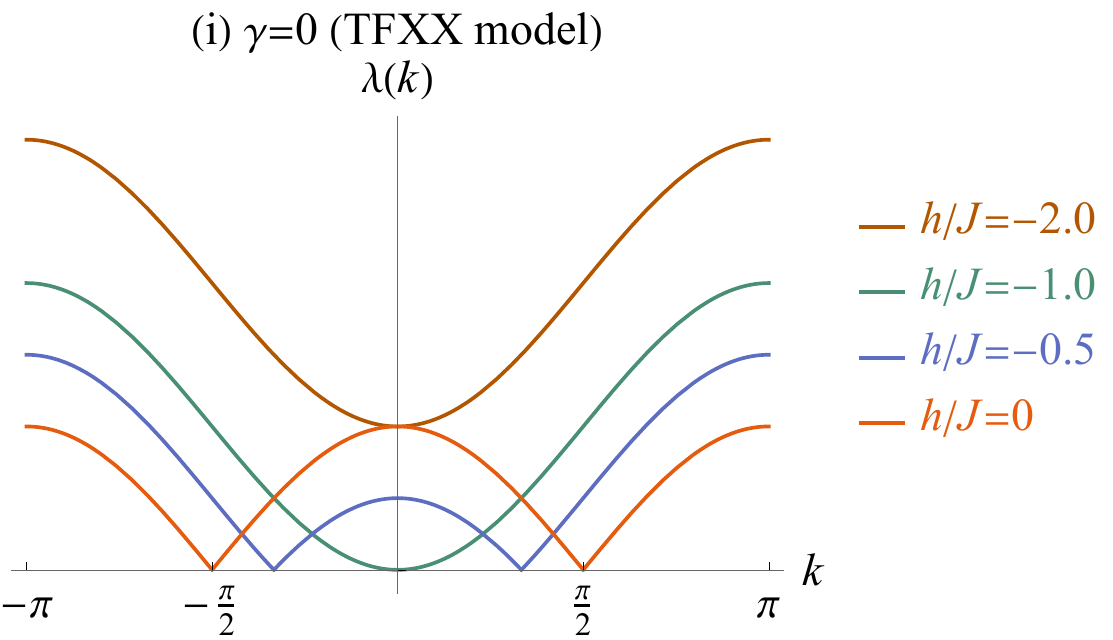} \includegraphics[clip,width=1.0\columnwidth]{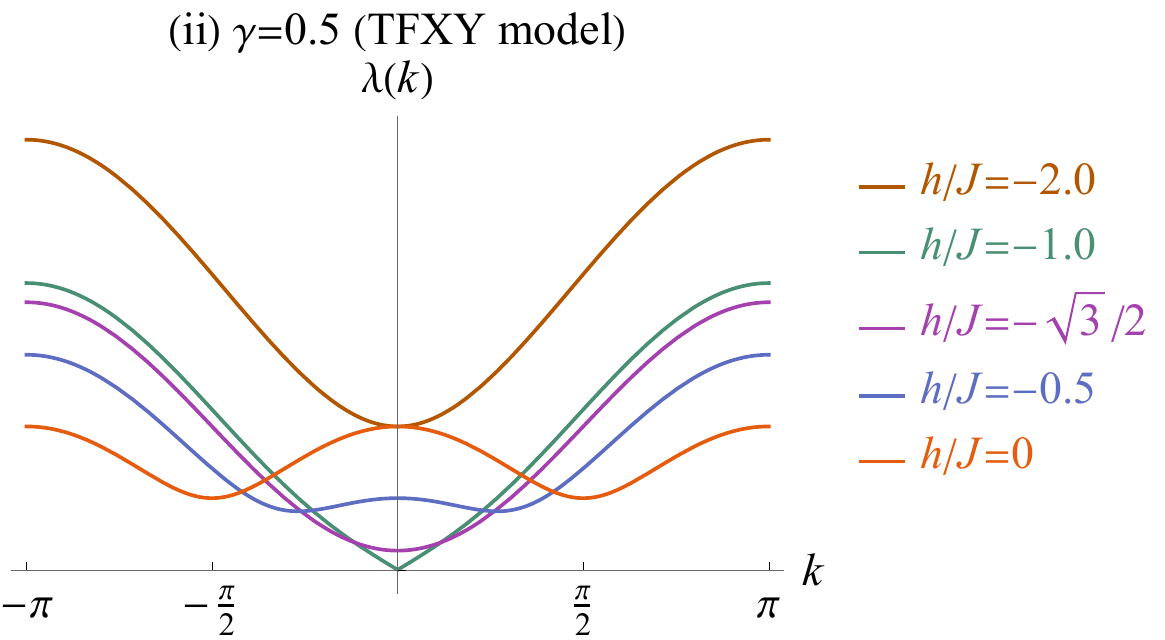}
    \includegraphics[clip,width=1.0\columnwidth]{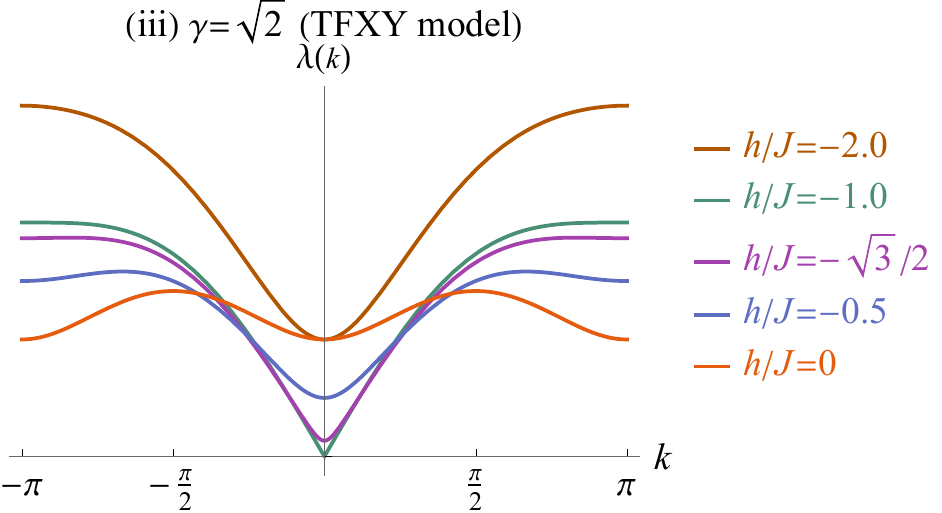}
    \end{minipage}
  \caption{Dispersion relations of fermions in TFXY models under some transverse fields $h$, based on Eq.~(\ref{eq:TFXY_lambda_bulk}).
  (i) $\gamma=0$ (the TFXX model).
(ii) $\gamma=0.5$.
(iii) $\gamma=\sqrt{2}$.
  The dispersion relations at $h>0$ are shifted by $\pi$, compared with those for $h<0$.
}
    \label{fig:TFXY_disp}
\end{figure}

In Fig.~\hyperref[fig:TFXY_disp]{\ref*{fig:TFXY_disp}(i)}, we show typical forms of dispersion relations at $\gamma=0$ (TFXX) and $\gamma=0.5$, where we compare several types of transverse fields
$h/J$.
In the TFXX chain, we can find that dispersion relations in the weak-field region $0\leq |h|/J<1.0$ have a gapless kink structure at zero energy, while it becomes a gapped dispersion in the strong-field region $1 \leq |h|/J$.
For the TFXY chain, we focus on $\gamma=0.5$ as an example, as shown in Fig.~\hyperref[fig:TFXY_disp]{\ref*{fig:TFXY_disp}(ii)}.
Then, dispersion relations in $0 \leq |h|/J < \sqrt{3}/2$ have two minima at nonzero momenta, while such a structure disappears in $ \sqrt{3}/2 \leq |h|/J$ and there is only one minimum.

\subsection{Open boundary}
Under the OBC, the pseudofermion Hamiltonian obtained after the JW transformation is
\begin{align}
H =& -\frac{J}{2}\sum_{i=1}^{N-1} \left[(c_i^\dagger c_{i+1} - c_i c_{i+1}^\dagger) + \gamma (c_i^\dagger c_{i+1}^\dagger - c_i c_{i+1}) \right] \notag\\
&-h\sum_{i=1}^{N} (c_i^\dagger c_i -\frac{1}{2}).
\end{align}
At $\gamma = 0$, this Hamiltonian is diagonalized after the Fourier transformation.
At $\gamma \neq 0$, due to the pairing terms, the Hamiltonian is not diagonalized after the Fourier transformation, and the Bogoliubov transformation is required.

\begin{figure}[hbt!]
    \centering
    \begin{minipage}[t]{1.0\columnwidth}
    \includegraphics[clip,width=1.0\columnwidth]{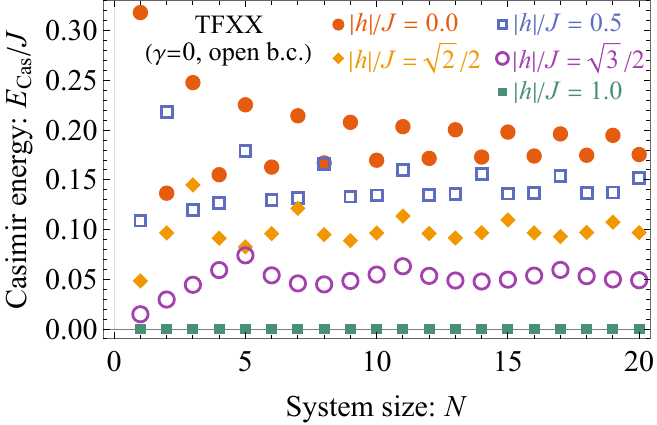}
    \end{minipage}
\caption{Casimir energy in the transverse-field XX chain under the open boundary condition.}
\label{fig:TFXX_open}
\end{figure}

\begin{figure}[hbt!]
    \centering
    \begin{minipage}[t]{1.0\columnwidth}
    \includegraphics[clip,width=1.0\columnwidth]{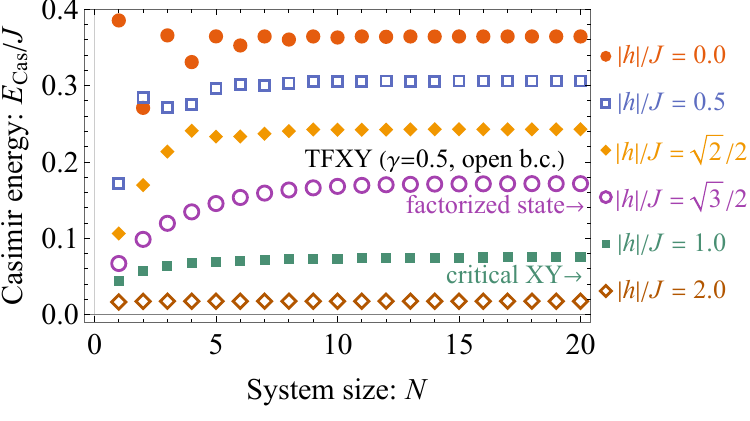}
    \end{minipage}
\caption{Casimir energy in the transverse-field XY chain at $\gamma=0.5$ under the open boundary condition.}
\label{fig:TFXY_open}
\end{figure}

Numerical results the Casimir energies for $\gamma=0$ and $0.5$ are shown in Figs.~\ref{fig:TFXX_open} and \ref{fig:TFXY_open}, respectively.

\subsubsection{Zero field: $h/J=0$}
In the zero- and weak-field region ($|h|/J<1$) of the XX chain, the system at low energies is in a gapless quantum phase characterized by a $c=1$ conformal field theory, which is the so-called Tomonaga-Luttinger liquid~\cite{Tomonaga:1950zz,Luttinger:1963zz,Haldane:1981zza,Haldane:1981zz}.

As shown in Fig.~\ref{fig:TFXX_open}, in the XX limit ($\gamma=0$), we find that the Casimir energy oscillates as a function of the system size $N$.
This oscillation is induced by the presence of Fermi points in the fermion dispersion relations, which is the same mechanism as the {\it oscillating Casimir effect} at finite fermion density~\cite{Fujii:2024fzy,Fujii:2024ixq,Fujii:2024woy,Fujii:2025qmf}.

The oscillation period is determined by the Fermi momentum of fermions, $ak_\mathrm{F}= \pm \arccos(-h/J)$:\footnote{We should note the sign of $h$.
For $h\leq0$, $N^\mathrm{osc} = \pi/|ak_\mathrm{F}| =\pi/\arccos(-h/J)$: the period is determined by the Fermi momentum itself.
For $h \geq 0$, $N^\mathrm{osc} = \pi/(\pi - |ak_\mathrm{F}|) =\pi/[\pi-\arccos(-h/J)]=\pi/\arccos(h/J)$: the period is determined by the Fermi momentum shifted by $\pi$.
}
\begin{align}
N^\mathrm{osc}
=\frac{\pi}{\arccos(|h|/J)}.
\end{align}
For example, in the vanishing field ($h/J=0$), the Fermi points are located at the Fermi momentum $ak_\mathrm{F}=\pm \pi/2$.
Then, we can find $N^\mathrm{osc}=2$, which is shown in Fig.~\ref{fig:TFXX_open}.
For $|h|/J>0$, the period becomes larger as $h$ increases.
For $|h|/J=1$, the period is infinite:  the oscillation does not occur.

Typical behaviors in the XY model (i.e., $0<\gamma <1$) are shown in Fig.~\ref{fig:TFXY_open}.
Here, the oscillation of the Casimir energy survives in the small size, whereas it is suppressed in the large size.
This suppression behavior can be interpreted from the dispersion relations shown in Fig.~\hyperref[fig:TFXY_disp]{\ref*{fig:TFXY_disp}(ii)}: in this region, there are two minima at nonzero momenta.
This is a remnant of the gapless kink structure seen in the XX limit ($\gamma \to 0$), but it is gapped.
Such a gapped structure restricts the energy shift due to the momentum discretization, and the amplitude of the resulting Casimir energy is oscillatory but suppressed in the large size.
This is a similar mechanism of the faster damping of Casimir energy for massive degrees of freedom.

\subsubsection{Weak-field region: $0<h^2/J^2+\gamma^2<1$}

In this region, we also find an oscillating Casimir effect.
For example, at $|h|/J=1/2$ in the XX chain, the Fermi points are located at the Fermi momentum $ak_\mathrm{F}= \pm \arccos(h/J)=\pm \pi/3$.
Then, we can find $N^\mathrm{osc}=3$.
Similarly, we find $N^\mathrm{osc}=4$ at $|h|/J=\sqrt{2}/2$ and  $N^\mathrm{osc}=6$ at $|h|/J=\sqrt{3}/2$.

In the XY chain, we can see an oscillating behavior (see $|h|/J=0.5$ and $\sqrt{2}/2$ in Fig.~\ref{fig:TFXY_open}), but its amplitude is damped in the larger size $N$: we may call this behavior the {\it damped oscillating Casimir effect}.

\subsubsection{Factorized state: $h^2/J^2+\gamma^2=1$}

The XY chain satisfying $h^2/J^2+\gamma^2=1$ is a special situation~\cite{Kurmann:1981,KURMANN1982235,PhysRevB.32.5845}, which is called the {\it factorized state}.
At the classical Ising point ($\gamma=1,h=0$), the ground state is fully factorized, corresponding to a classical spin configuration aligned along the $x$-axis without quantum entanglement.
As $\gamma$ decreases from 1 to 0, the system enters an intermediate XY regime where quantum fluctuations become increasingly significant due to competing 
spin interactions along the $x$ and $y$ directions.
Along the factorization line defined by $h^2/J^2+\gamma^2=1$, the ground state remains exactly factorized.
At the XX limit ($\gamma=0,h=J$), the ground state is still factorized, but with spins oriented at a nontrivial angle set by the transverse field, where the dispersion relations are exactly a shifted cosine band.

The dispersion relations for factorized states in the XY chains are written as a function of $h$ (or alternatively $\gamma$):
\begin{align}
\lambda_k =|J+h\cos{k}|.
\end{align}
This form is reduced to the flat band $\lambda_k = J$ in the classical Ising limit ($h=0$) or the shifted cosine band $\lambda_k = J|1+\cos{k}|$ in the critical XX limit ($|h|/J=1$).
In particular, for $0<h<1$, the dispersion relations can be interpreted as a shifted and scaled cosine band, where ``scaled" refers to the energy scaling of the band.
Thus, the Casimir effect in this region exhibits features of a shifted and scaled cosine band.

In Fig.~\ref{fig:TFXY_open}, we find that the Casimir energy depends on $N$ but does not oscillate.
This fact can be understood through the structure of dispersion relations: the minimum in the dispersion relation is located at $k=0$.
Since the oscillation of Casimir energy is induced by nonzero-momentum Fermi points, it disappears when the minimum is located at $k=0$.
In addition, a nonzero $N$ dependence of the Casimir energy is induced by a surface effect under the OBC.
We will discuss that this situation changes under the PBC.

\subsubsection{Intermediate-field region: $1<h^2/J^2+\gamma^2$, $|h|/J<1$}

In the region satisfying both $1<h^2/J^2+\gamma^2$ and $|h|/J<1$ of the XY chain, similar to the factorized state, the Casimir energy does not oscillate.
This is because there is no nonzero-momentum minimum in the dispersion relation.

\subsubsection{Critical field: $|h|/J=1$}

In the XX model, the Casimir energy is exactly zero at any $N$.
This is because the dispersion relation is exactly a shifted cosine band, and the energy shift due to the momentum discretization is exactly canceled.
Such a {\it vanishing or zero Casimir effect} is known to hold in both the continuum and lattice field theories~\cite{Nakayama:2022ild}.

Contrary to the XX chain, in the critical XY chain, the Casimir energy depends on $N$.
The corresponding dispersion relation has a kink structure at zero-momentum, as shown in Fig.~\hyperref[fig:TFXY_disp]{\ref*{fig:TFXY_disp}(ii)}.

\subsubsection{Strong-field region: $|h|/J>1$}
In the XX chain, the Casimir energy remains exactly zero for any $N$ because the dispersion relation remains a shifted cosine band. 
On the other hand, in the XY chain, the Casimir energy still depends on $N$ but is dominated by the surface energy.
In the limit of $|h| \to \infty$, it approaches zero.

\subsection{Periodic boundary}
\begin{figure}[tb!]
    \centering
    \begin{minipage}[t]{1.0\columnwidth}
    \includegraphics[clip,width=1.0\columnwidth]{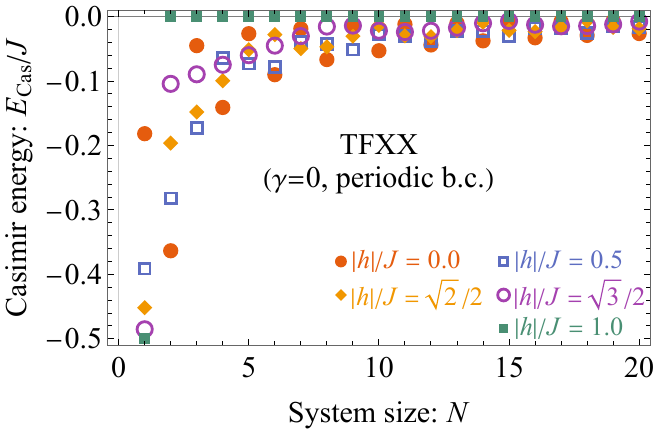}
    \end{minipage}
\caption{Casimir energy in the transverse-field XX chain under the periodic boundary condition.}
\label{fig:TFXX_pbc}
\end{figure}

Under the PBC, the pseudofermion Hamiltonian after the JW transformation is
\begin{align}
H =& -\frac{J}{2}\sum_{i=1}^{N-1} \left[(c_i^\dagger c_{i+1} - c_i c_{i+1}^\dagger) + \gamma (c_i^\dagger c_{i+1}^\dagger - c_i c_{i+1}) \right] \notag\\
&+ (-1)^{N_c} \frac{J}{2}\left[(c_N^\dagger c_{1} - c_N c_{1}^\dagger) + \gamma (c_N^\dagger c_{1}^\dagger - c_N c_{1}) \right] \notag\\
&-h\sum_{i=1}^{N} (c_i^\dagger c_i -\frac{1}{2}).
\end{align}
Numerical results for $\gamma=0$ and $0.5$ are shown in Figs.~\ref{fig:TFXX_pbc} and \ref{fig:TFXY_pbc}, respectively.

\subsubsection{Zero field: $h/J=0$}

When the transverse field is zero ($h/J=0$), we find the oscillation of the Casimir energy, where the period is $N^\mathrm{osc}=2$ similar to the case of OBC.

\begin{figure}[tb!]
    \centering
    \begin{minipage}[t]{1.0\columnwidth}
    \includegraphics[clip,width=1.0\columnwidth]{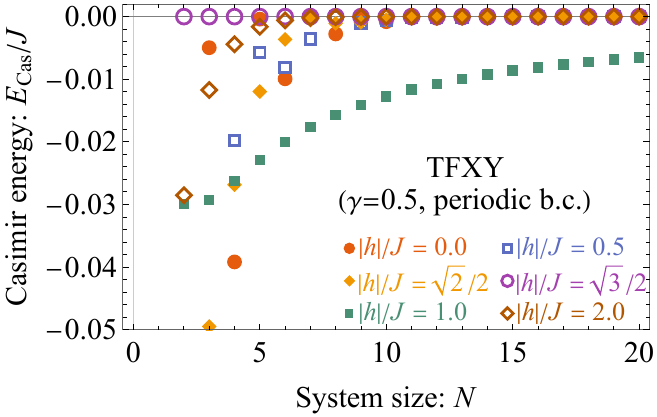}
    \end{minipage}
\caption{Casimir energy in the transverse-field XY chain at $\gamma=0.5$ under the periodic boundary condition.}
\label{fig:TFXY_pbc}
\end{figure}

In the XY chain, as shown in Fig.~\ref{fig:TFXY_pbc}, the Casimir energy in the large size is damped.
These behaviors are similar to those in the case of the OBC, except the absence of the surface energy.

\subsubsection{Weak-field region: $0<h^2/J^2+\gamma^2<1$}

Similar to $h=0$, in the weak-field region of the XX chain, we find oscillations of Casimir energy.
We note that this region  (i.e, $0\leq|h|/J<1$ under the PBC) can be regarded as a regime of a purely oscillating effect, because of the absence of surface energy and damping behavior.

In the case of XY chain, the Casimir energy oscillates and is damped in the large size, which is similar to the behavior in the OBC.
However, we find that this oscillation is distorted by an additional effect.
In the ordered phase of XY model (precisely, $0<|\gamma|/J<1$) under PBCs, the energy difference between the odd- and even-parity sectors switches as a function of the system size $N$.
In this regime, the resulting ground-state parity switching is reflected in the Casimir energy.

\begin{figure}[tb!]
    \centering
    \begin{minipage}[t]{1.0\columnwidth}
    \includegraphics[clip,width=1.0\columnwidth]{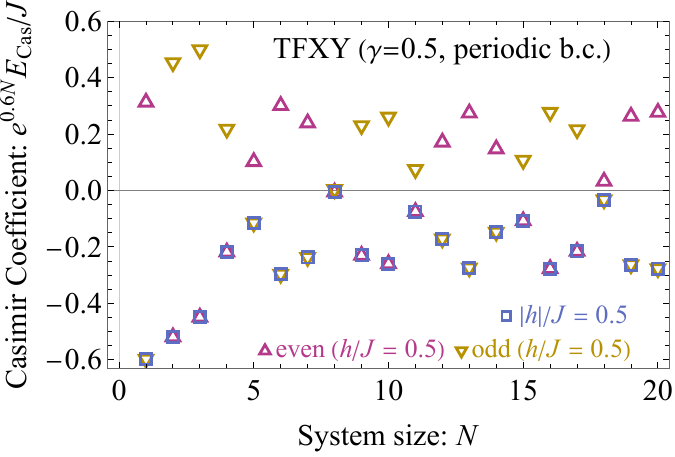}
    \end{minipage}
\caption{The comparison between the ground-state energies in the odd and even sectors for $h/J=0.5$ in the XY chain at $\gamma=0.5$ under the PBC, where the Casimir energy is multiplied by a factor $e^{0.6N}$ for clearly visualizing the oscillations.
The true Casimir energy is chosen as the lower one.
Note that, at $h/J=-0.5$, the Casimir energy is the same as that at $h/J=0.5$, but the distribution of odd and even sectors becomes different.
}
\label{fig:TFXY_pbc_sector}
\end{figure}

To visualize the competition between the two ground states,  we compare the Casimir energies calculated from two sectors, in Fig.~\ref{fig:TFXY_pbc_sector}.
Each ground-state energy (or Casimir energy) of the even or odd sector oscillates as a function of $N$, while the true ground state (or the true Casimir energy) is determined from the lower-energy parity sector at each $N$.
For this reason, the resulting Casimir energy is modified by switching of two types of oscillations.
This behavior may be called the {\it ground-state switching Casimir effect}, or more precisely, the oscillating Casimir effect modified by the ground-state switching.

\subsubsection{Factorized state: $h^2/J^2+\gamma^2=1$}

In the region satisfying $h^2/J^2+\gamma^2=1$, the ground state is the factorized state.
In the XX model, we find the realization of the remnant Casimir effect because of the shifted cosine band under PBCs.

In the XY chain, we can find that the Casimir energy behaves as a remnant Casimir effect, which is distinct from that under the OBC, where the $N$ dependence of Casimir energy survives.
This difference between the PBC and OBC is due to the shifted and scaled cosine band.
Intuitively, in factorized states, each spin is independent of the others, so that the ground-state energy is unaffected by finite-size effects under PBCs.
As a result, the Casimir energy vanishes.
On the other hand, under the OBC, the surface effect introduces an $N$ dependence in the ground-state energy.

\subsubsection{Intermediate-field region: $1<h^2/J^2+\gamma^2$, $|h|/J<1$}

In the intermediate-field region satisfying both $1<h^2/J^2+\gamma^2$ and $|h|/J<1$, we find no oscillation of Casimir energy.
One reason is that there is no nonzero-momentum minimum in the dispersion relations.

\subsubsection{Critical point: $|h|/J=1$}

At the critical field $|h|/J=1$ of XX chain, we find the remnant Casimir effect.
This is because the dispersion relation is the shifted cosine band.

In the critical field of XY model, we find the $1/N$ scaling of Casimir energy in the larger size.
This behavior can be regarded as the Casimir effect for massless degrees of freedom.

The zero-point energy at the critical point is analytically known. 
For $\gamma \neq 0$, the zero-point energies are~\cite{Henkel:1986ea}
\begin{align}
\frac{E_0^\mathrm{PB}}{J}
&=\frac{E_0^\mathrm{bulk}}{J} -\frac{\pi}{12N}  +\frac{7\pi^3}{960N^3}\left( \frac{1}{\gamma^2} -\frac{4}{3} \right)  + \mathcal{O} (1/N^5), \\
\frac{E_0^\mathrm{APB}}{J}
&=\frac{E_0^\mathrm{bulk}}{J} +\frac{\pi}{6N}  -\frac{\pi^3}{120N^3}\left( \frac{1}{\gamma^2} -\frac{4}{3} \right)  + \mathcal{O} (1/N^5),
\end{align}
By subtracting the bulk contributions,
\begin{align}
\frac{E_\mathrm{Cas}^\mathrm{PB}}{J}
&= -\frac{\pi}{12N}  +\frac{7\pi^3}{960N^3}\left( \frac{1}{\gamma^2} -\frac{4}{3} \right)  + \mathcal{O} (1/N^5), \\
\frac{E_\mathrm{Cas}^\mathrm{APB}}{J}
&=\frac{\pi}{6N}  -\frac{\pi^3}{120N^3}\left( \frac{1}{\gamma^2} -\frac{4}{3} \right)  + \mathcal{O} (1/N^5),
\end{align}
Since the coefficients of the first terms are universal, they coincide with those obtained in the critical Ising chain.
This explains why the $1/N$ scaling of Casimir energy is realized even in the XY region.
On the other hand, the higher-order correction, such as the $1/N^3$ term, differ slightly from those in the critical Ising chain.
This difference is reflected in the small-size region.

\subsubsection{Strong-field region: $|h|/J>1$}

Also in the strong-field region, the Casimir energy does not oscillate.
In the XX chain, its behavior is exactly the remnant Casimir effect due to a shifted cosine band, which is the same as the result at the critical point $|h|/J=1$.

In the XY chain, the Casimir energy is nonzero and non-oscillating.
In the strong-field limit, it approaches the remnant Casimir effect.

\begin{figure}[tb!]
    \centering
    \begin{minipage}[t]{1.0\columnwidth}
    \includegraphics[clip,width=1.0\columnwidth]{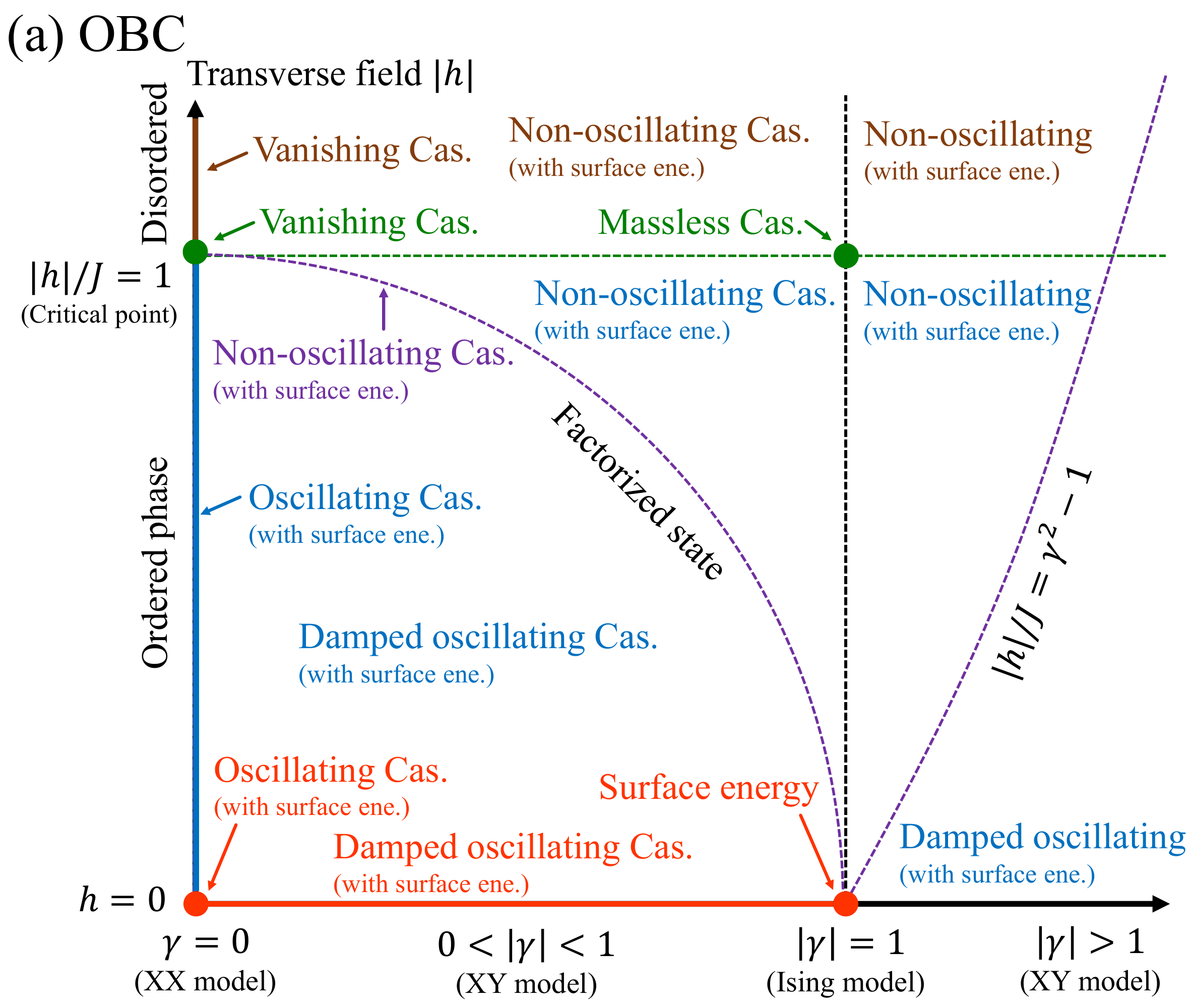}
    \includegraphics[clip,width=1.0\columnwidth]{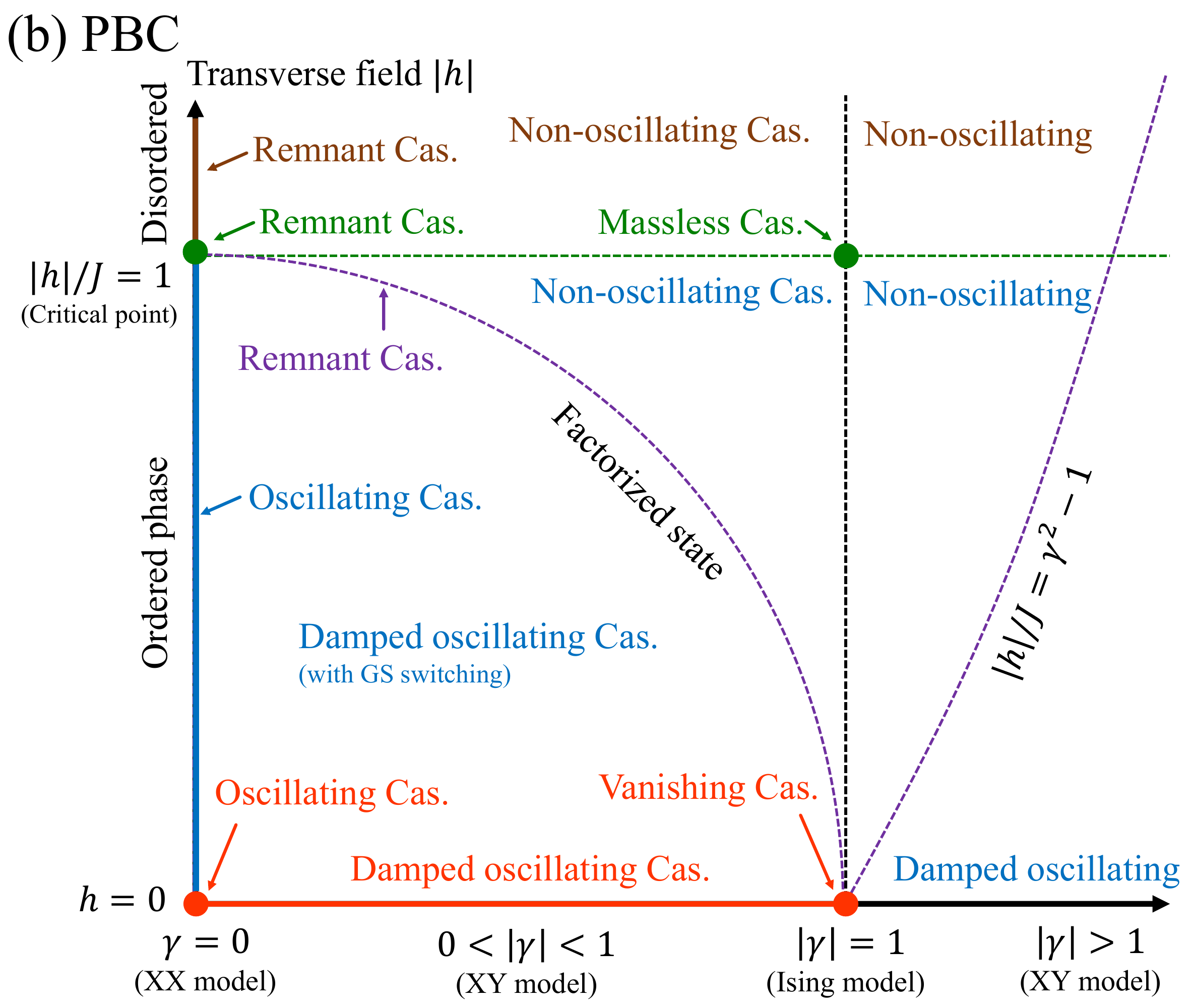}
    \end{minipage}
\caption{Phase diagrams of Casimir effect in the one-dimensional transverse-field XY model.
(a) the open boundary condition (OBC). (b) the periodic boundary condition (PBC).}
\label{fig:TFXY_diagram}
\end{figure}

\subsection{Short summary}
Before concluding this section, we mention the region of $\gamma>1$ (i.e., the strongly anisotropic region).
The behavior of Casimir energy is distinguished by the line along $|h|/J=\gamma^2-1$.
In the weak-field region of $|h|/J<\gamma^2-1$, we find the oscillating Casimir effect.
In the strong-field region of $|h|/J>\gamma^2-1$, the Casimir energy does not oscillate.
This can be understood by the dispersion relations as shown in Fig.~\hyperref[fig:TFXY_disp]{\ref*{fig:TFXY_disp}(iii)}, where the examples of dispersion relations at $\gamma=\sqrt{2}$ are shown.
The dispersion relations in the weak-field region have two maxima around $k=\pm \pi/2$.
These maxima can lead to the damped oscillating Casimir energy with a period $N^\mathrm{osc} \sim 2$.
In the strong-field region, the maximum moves to $k=\pi$, which does not induce oscillating behaviors.

Finally, in Fig.~\ref{fig:TFXY_diagram}, we show the phase diagrams of the Casimir effect in the one-dimensional transverse-field XY model, including the Ising and XX chains.
The results under the PBC and OBC are almost similar, but one can also recognize several differences, such as (i) the presence of the surface energy in the OBCs, (ii) the ground-state switching between the even and odd sectors in the PBCs, and (iii) the remnant Casimir effect in the PBCs.

\section{Conclusion}  \label{Sec:4}
We have shown the relationship between the Casimir effects and finite-size corrections to the ground-state energy of the transverse-field Ising or XY chains.
Here, the ``Casimir effect" refers to the ground-state energy shift induced by the finite-size effect, which can also be interpreted as a fermionic Casimir effect via the JW transformation.
Therefore, our finding emphasizes that a fermionic Casimir effect can be realized as a mathematically equivalent form in spin chains.
Furthermore, we naturally conclude that {\it the study of the fermionic Casimir effect is not merely a theoretical effort but is experimentally feasible}
(e.g., in spin-chain-like systems such as CoNb$_2$O$_6$~\cite{Coldea:2010zz}, trapped-ion or superconducting quantum simulators, and ultracold atoms in optical lattices).

In addition, a rich variety of Casimir effects that arise depending on the model parameters are summarized in Figs.~\ref{fig:TFI_diagram} and \ref{fig:TFXY_diagram}.
Spin-chain systems therefore provide a platform for testing these effects and confirm that they are experimentally measurable phenomena rather than purely theoretical ideas.

In this paper, we have focused only on Ising and XY chains, where the interpretations of finite-size scaling can be systematically established using fermionic Casimir effects since analytically diagonalizable fermionic Hamiltonians are obtained via the JW transformation.
In contrast, for the XXZ and Heisenberg chains, the JW transformation does not yield a directly diagonalizable fermionic Hamiltonian, as interaction terms appear.
Nevertheless, the fermionic representation remains well-defined and useful for interpreting physical quantities from a fermionic perspective.
Our results can be extended to more nontrivial one-dimensional spin models and may also provide insight into higher-dimensional systems.

\section*{ACKNOWLEDGMENTS}
This work was supported by the Japan Society for the Promotion of Science (JSPS) KAKENHI (Grants No. JP24K07034 and JP24K17059).

\appendix
\section{Details of Numerical Methods} \label{App:numerical} 
In this appendix, we demonstrate the numerical methods to calculate the Casimir energies shown in the main text.

The definition of the Casimir energy is given as $E_\mathrm{Cas}=E_0-E_0^\mathrm{bulk}$ in Eq.~(\ref{eq:ECas_def}).
$E_0^\mathrm{bulk}$ is obtained from the numerical integral in Eq.~(\ref{eq:TFIM_E0bulk}).
$E_0$ is obtained from the three methods (i) the numerical sum over discrete energies in Eq.~(\ref{eq:TFIM_E0open}), (ii) the exact diagonalization, and (iii) the DMRG.
For the DMRG, we perform 50 DMRG sweeps with a maximum bond dimension of 100, where we have checked only the case of OBC.

All the results from the three methods agree well.
In other words, the Casimir energy can be computed correctly both spin systems (i.e., the exact diagonalization or DMRG) and fermion systems [i.e, Eq.~(\ref{eq:TFIM_E0open})] after the JW transformation.

\bibliography{ref}

\end{document}